\documentclass[prl,twocolumn,amsmath,amssymb]{revtex4-2}

\usepackage{amsfonts,amsmath,amssymb,graphicx}
\usepackage[colorlinks=true,allcolors=blue]{hyperref}

\allowdisplaybreaks

\newcommand{\arXiv}[1]{\href{http://www.arXiv.org/abs/#1}{arXiv:#1}}
\renewcommand{\doi}[2]{\href{http://dx.doi.org/#2}{#1}}
\newcommand{\beq}{\begin{equation}}
\newcommand{\eeq}{\end{equation}}
\newcommand{\nn}{\nonumber}
\newcommand{\de}{\delta}
\newcommand{\ER}{Erd\H{o}s-R\'enyi }
\newcommand{\ea}[1]{\left\langle{#1}\right\rangle}
\DeclareMathOperator{\Var}{Var}

\begin{document}

\title{Ensemble inequivalence and phase transitions in unlabeled networks}

\author{Oleg Evnin}
\affiliation{High Energy Physics Research Unit, Faculty of Science, Chulalongkorn University, Bangkok, Thailand \&\\
Theoretische Natuurkunde, Vrije Universiteit Brussel \& International Solvay Institutes, Brussels, Belgium}
\author{Dmitri Krioukov}
\affiliation{Department of Physics \& Network Science Institute \&\\ Department of Mathematics \& Department of Electrical and Computer Engineering, Northeastern University, Boston, Massachusetts, USA}

\begin{abstract}
We discover a first-order phase transition in the canonical ensemble of random unlabeled networks with a prescribed average number of links. The transition is caused by the nonconcavity of microcanonical entropy. Above the critical point coinciding with the graph symmetry phase transition, the canonical and microcanonical ensembles are equivalent and have a well-behaved thermodynamic limit. Below the critical point, the ensemble equivalence is broken, and the canonical ensemble is a mixture of phases: empty networks and networks with average degrees diverging logarithmically with the network size. As a consequence, networks with bounded average degrees do not survive in the thermodynamic limit, decaying into the empty phase. The celebrated percolation transition in labeled networks is thus absent in unlabeled networks. In view of these differences between labeled and unlabeled ensembles, the question of which one should be used as a null model of different real-world networks cannot be ignored.

\end{abstract}

\maketitle

The \ER (ER) model~\cite{erdos1959random} is truly the ``harmonic oscillator'' of network science and random graphs. It appears as the first prominent subject in many recognizable textbooks in these areas~\cite{barabasi2016network,newman2018networks,bollobas1985random,janson2000random,hofstad2016random}, which is not surprising since it has been used as a Petri dish to study a broad spectrum of phenomena in statistical physics, network science, social science, computer science, graph theory, probability, statistics, and many other disciplines.
In the microcanonical ER ensemble, $m$ links connect random pairs of $n$ distinct labeled nodes, while in the canonical ensemble, every pair of those nodes is linked independently with probability $p$. The result in both cases is the maximum-entropy distribution over the space of labeled networks with $n$ nodes and a given exact number of links~$m$, or expected number of links~$\ea{m}=pn(n-1)/2$~\cite{park2004statistical,contraction}.
Of particular interest to statistical physics are the ensemble equivalence and critical phenomena, which have been explored at great depths in the ER model~\cite{anand2010gibbs,squartini2015breaking,ER1,aldous1997brownian,ER2,cameron2013random}. A succinct summary is that the percolation transition is at the average degree~${k}_p\equiv2m/n=1$, the network connectivity and symmetry transition~\cite{symmetry} is at ${k}_c=\log n$, and the canonical and microcanonical ER ensembles are equivalent for any~$p\in[0,1]$ and $m=\ea{m}$, according to all definitions of ensemble equivalence~\cite{touchette}.

The unlabeled version of the ER model, which is the same maximum-entropy distribution, except over the space of networks whose vertices are indistinguishable~\cite{Luczak}, has received much less attention, even though it is relevant for real-world networks with unlabeled nodes such as molecules and atoms in physical and material networks~\cite{glover2024measuring,dehmamy2018structural,posfai2023impact,liu2020isotopy,pete2024physical,papadopoulos2018network,nauer2021random,quantum}. The slow progress here is prima\-rily due to the difficulties of dealing with unlabeled networks~\cite{wormald,canonical}, and to the established fact that the labeled and unlabeled ER models are essentially equivalent, upon a certain transformation, as soon as the average degree ${k}\gg\log n$~\cite{Luczak}. However, most real-world networks are sparse with bounded average degree~\cite{barabasi2016network,newman2018networks}, and this sparse regime has not received the deserved attention.

Here, we show that the unlabeled \ER model is entirely different from its well-studied labeled counterpart in the sparse regime with average degree ${k}\ll\log n$. The main cause of this difference is the nonconcavity of entropy, a well-studied origin of ensemble inequivalence~\cite{touchette} and phase transitions~\cite{nonconc,gross2001microcanonical,horvat2015reducing,thesis}.
In contrast with the standard labeled ER model, the microcanonical unlabeled ER entropy is not concave; it has an inflection point at ${k}_c\approx\log n$. We show that this nonconcavity causes a first-order phase transition at $k=k_c$. In the dense regime ${k}\gg{k}_c$, the microcanonical and canonical unlabeled ER models are equivalent, while away from that region, this ensemble equivalence is broken. The canonical networks with expected average degree ${k}\ll{k}_c$ are just empty with high probability $P\approx 1-{{k}}/2{k}_c$, while with probability $1-P$, they have average degree~$\approx2{k}_c$. This degeneracy is related to but different from the one in microcanonical networks, which are known to exhibit an extreme form of phase separation at ${k}\ll{k}_c$~\cite{Luczak}: most nodes are isolated (have degree~$0$), while all the links condense within an infinitesimal connected component whose size and average degree are $\approx n{k}/{k}_c$ and $\approx{k}_c$, respectively. Since $P\to1$ if ${k}\ll{k}_c$, unlabeled canonical ER networks with expected average degrees between~$0$ and~${k}_c$ are in fact not realizable in the thermodynamic limit. A remarkable consequence of this is that the celebrated percolation transition at ${k}_p=1$ is entirely absent in the unlabeled \ER model.

{\it Microcanonical entropy.}--- To obtain these results, we first show that the unlabeled microcanonical ER entropy is not concave. This entropy is $S_{nm}=\log g_{nm}$, where $g_{nm}$ is the number of unlabeled graphs with $n$ nodes and $m$ links~\cite{OEIS}. In the case of labeled graphs, the corresponding number is simply~${N \choose m}$ with $N\equiv n(n-1)/2$, and the entropy is everywhere concave. The combinatorics of unlabeled graphs is considerably more intricate, since one unlabeled graph corresponds to a class of isomorphic labeled graphs, i.e., all labeled graphs related to each other by vertex renumbering~\cite{geo}. The standard solution proceeds via P\'olya counting theory~\cite{HP}, and we provide a sketch of the derivation in the Supplemental Material \cite{quantum}.
The generating function for the number of unlabeled graphs, defined as
\beq
g_n(x)\equiv\sum_{m=0}^N g_{nm}x^m,\quad N\equiv{n\choose2},
\eeq
is given by
\begin{align}
g_n(x)=&\sum_{\{j_k\}}\Big(\prod_k k^{j_k}j_k!\Big)^{\hspace{-1mm}-1}\prod_{r<s} (1+x^{\mathrm{LCM}(r,s)})^{\mathrm{GCD}(r,s)j_rj_s}\nn\\
&\times\prod_k \Big\{(1+x^{2k+1})^{kj_{2k+1}}(1+x^{2k})^{(k-1)j_{2k}}\label{Harary}\\[-1.5mm]
&\hspace{1.3cm}\times (1+x^k)^{j_{2k}+kj_k(j_k-1)/2}\Big\},\nn
\end{align}
where $\mathrm{LCM}(r,s)$ and $\mathrm{GCD}(r,s)$ are the least common multiple and the greatest common divisor of $r$ and $s$, while $\sum_{\{j_k\}}$ stands for summation over all sequences of nonnegative integers $\{j_k\}$ such that $\sum_{k=1}^nkj_k=n$~\cite{num}. We then employ representation~(\ref{Harary}) to compute $g_{nm}$ numerically, and use these exact numerical results in all the figures in this paper~\cite{share}. Yet for analytical purposes, we cannot work with the exact expression~(\ref{Harary}) directly as it is forbiddingly complicated. Instead, we will rely upon its asymptotics that we discuss next.

{\it Entropy asymptotics and nonconcavity.}--- The asymptotics for $g_{nm}$ have been worked out in \cite{Wright}, and we summarize the relevant ones below. Notationwise,
if the symbol `$\star$' in $a_n \star b_n$ is `$\ll$', `$\sim$', `$\approx$', or `$\gg$', it means that $\lim_{n\to\infty} a_n/b_n=c\geq0$, and that $c=0$, $0<c<\infty$, $c=1$, and $c=\infty$, respectively.

Let ${k}\equiv2m/n$ be the average degree, $C>1$ an arbitrary constant, and $\mu_m$ the solution of the equation
\beq
\mu_m\log\mu_m = 2m,
\eeq
i.e., $\mu_m=2m/W_0(2m)$, where $W_0(x)$ is the principal branch of the Lambert $W$ function. Then,
\begin{align}
 &g_{nm}          \approx \frac{1}{n!}{N \choose m}\qquad\mbox{if } {k}-\log n\gg1,\label{bell}\\
 &\frac{g_{n,m+1}}{g_{nm}} \approx \frac{n^2}{2m} \qquad\mbox{if } \log n \leq {k} \leq C \log n,\label{lesssparse}\\
 &\frac{g_{n,m+1}}{g_{nm}} \approx \frac{\mu_m^2}{2m}=\frac{\mu_m}{\log\mu_m} \qquad\mbox{if } {k}\leq\log n.\label{wrightsparse}
\end{align}
Observe that \eqref{bell} implies that if ${k}-\log n\gg1$, then the number of unlabeled graphs is the number of labeled graphs divided by $n!$. This is consistent with the known fact that labeled ER graphs are asymmetric~\cite{symmetry} if ${k}\gg\log n$, identifying $k_c=\log n$ as the {\it symmetry transition} point. There exist explicit asymptotics for $g_{nm}$ in the sparse regime ${k}\leq\log n$ as well~\cite{Wright}, but we will not need that. Instead, we will rely on \eqref{wrightsparse} in what follows.

\begin{figure}
\centerline{\includegraphics[width=3.45in]{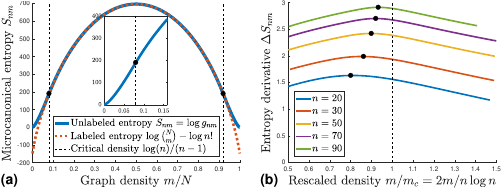}}
  \caption{{\it Nonconcavity of the microcanonical entropy:} \textbf{(a)}~The microcanonical entropy $S_{nm}$ of the unlabeled \ER model is shown as a function of the graph density $m/N$ for $n=50$, juxtaposed against the microcanonical entropy of the labeled \ER model, pulled down by $\log n!$. The inset zooms onto the critical point at $m_c/N=\log (n)/(n-1)$, where the unlabeled entropy derivative is near its maximum. \textbf{(b)}~The discrete derivatives $\Delta S_{nm}=S_{n,m+1}-S_{nm}$ of unlabeled entropy are shown as functions of the graph density rescaled by the critical value $m/m_c$ for different values of~$n$. The dots are the numerical maxima of the derivatives corresponding to the entropy inflection points.\vspace{-3mm}}
  \label{figmicroen}
\end{figure}

We first observe that \eqref{wrightsparse} implies that the microcanonical entropy $S_{nm}$ is not a concave function of number of edges $m$ in the sparse regime ${k}\leq\log n$.
Indeed, using~\eqref{wrightsparse},
\beq\label{nonconcave}
\frac{g_{nm}/g_{n,m-1}}{g_{n,m+1}/g_{nm}}\approx\frac{\mu_{m-1}/\log\mu_{m-1}}{\mu_{m}/\log\mu_{m}}<1,
\eeq
since $\mu_m/\log\mu_m$ is a monotonically growing function at $m>1$. The last equation is equivalent to $\Delta S_{n,m-1} < \Delta S_{nm}$, where $\Delta S_{nm}\equiv S_{n,m+1}-S_{nm}$, so the second derivative of the entropy is positive, or more explicitly, $2S_{nm}<S_{n,m-1}+S_{n,m+1}$, violating concavity. This is confirmed by the exact numerics in Fig.~\ref{figmicroen}, where we also show that the inflection point is at ${k}_c\approx\log n$, and that for ${k}\geq\log n$, the entropy is concave, which is consistent with the asymptotics~(\ref{bell},\ref{lesssparse}).

{\it Broken ensemble equivalence.}--- We next show how the entropy nonconcavity breaks the equivalence between the microcanonical and canonical ensembles at ${k}\ll\log n$. The canonical ensemble of unlabeled ER graphs~\cite{unlabeled} is defined by the standard entropy-maximizing Gibbs probability distribution for a random graph of size $n$ to have $m$ edges,
\beq\label{canon}
p_{nm}=\frac{g_{nm}q^{m}}{Z_n},\quad Z_n\equiv \sum_{m=0}^{N} g_{nm}q^m,\quad q\equiv e^{-\beta},
\eeq
where the inverse temperature~$\beta$ is the only parameter of the model, and $q$ is the Boltzmann factor. All the $g_{nm}$ graphs with $m$ edges are equiprobable, occurring with probability $p_{nm}/g_{nm}$. The key question is whether the distribution (\ref{canon}) is dominated by a single maximum at some $m=m_*$, or whether there are competing contributions coming from widely separated values of~$m$.

To answer this question, we first observe that $p_{nm}$ has a local maximum at $m=0$ for any $q<1$ since $g_{n0}=g_{n1}=1$. (There is only one empty and one unlabeled graph with $1$ edge.) The value $q=1$ corresponds to the graph density $m/N = 1/2$, while sparser graphs have $q<1$ \cite{unlabeled}.

Next, we show that besides this maximum, $p_{nm}$ does not have any maxima in the region ${k}<\log n$. Indeed, if $p_{nm}$ attains a maximum at $m_*$,
\beq\label{extr}
p_{nm_*}\ge p_{n,m_*-1},\qquad p_{nm_*}\ge p_{n,m_*+1},
\eeq
then in view of~\eqref{canon}, this implies that
\beq
g_{nm_*}^2\ge g_{n,m_*-1}\,g_{n,m_*+1},
\eeq
which contradicts the nonconcavity of entropy~\eqref{nonconcave}.

We thus see right away that the ensemble equivalence must be broken if the canonical ensemble is asked to produce graphs with average degree $\ea{k}<\log n$. Indeed, since $p_{nm}$ does not have maxima at ${k}<\log n$ other than at ${k}=0$, the contributions of probability mass yielding $\ea{k}$ cannot come from a single maximum of $p_{nm}$ in that region. They can come from two maxima, one at ${k}=0$ and another one at some ${k}_*>\log n$, and we show below that this is indeed what happens. But then, since the $p_{nm}$ distribution is not dominated by a single maximum at $\ea{k}<\log n$, the canonical ensemble with expected average degree $\ea{k}$ cannot be approximated by the microcanonical ensemble with the average degree fixed exactly at $\ea{k}$.

To show this explicitly, we need to analyze the maximum defined by (\ref{extr}). By the argument above, this maximum must be in the region ${k}>\log n$, where the asymptotics (\ref{bell},\ref{lesssparse}) apply. Using (\ref{lesssparse},\ref{canon},\ref{extr}) in the regime $m,n\gg 1$ and $m\ll N$, we get
\beq\label{expbe}
q\approx 2m_*/n^2\ll 1,
\eeq
which is the same relation as in the labeled ER ensemble~\cite{park2004statistical}, i.e., the Boltzmann factor is approximately the graph density $m_*/N\approx 2m_*/n^2$.
We then use Stirling's approximation in (\ref{bell},\ref{canon},\ref{expbe}) to construct the estimate
\beq\label{pmstar}
p_{nm_*}\equiv\frac{g_{nm_*}q^{m_*}}{Z_n}\approx \frac{e^{m_*}}{n!\,Z_n\sqrt{2\pi m_*}}.
\eeq
The values of $p_{nm}$ away from the two maxima, one at $m=m_*$ and the one at $m=0$, must be subleading, and cannot contribute significantly to the expected values of thermodynamic observables. Yet which of the two maxima dominates, and by how much, needs some further analysis~\cite{labelmax}.

To estimate the contributions from the two maxima, we first consider the partition function $Z_n$. The contribution it receives from the neighborhood of $m=0$ is well approximated by only the $m=0$ term in~\eqref{canon}, which is~$1$, neglecting all other $m>0$ terms, since the contribution from even the $m=1$ term $g_{n1}e^{-\beta}=q$ is already negligible thanks to~(\ref{expbe}). The contribution from the neighborhood of $m=m_*$ requires a bit more work, since $g_{nm}$ varies more gently in that region.
Using $1+y/m_*\approx(1+1/m_*)^y$ and $(1+1/m_*)^{x(x-1)/2}\approx e^{-x^2/2m_*}$ since $m_*\gg1$, we get from (\ref{lesssparse},\ref{expbe}) that 
\begin{align}
g_{n,m_*+x}q^{m_*+x}&\approx\frac{m_*^x \,g_{nm_*}q^{m_*}}{m_*(m_*+1)\cdots(m_*+x-1)}\nn\\
&\approx g_{nm_*}q^{m_*}e^{-x^2/2m_*},\label{gauss}
\end{align}
resulting in significant contributions from $|x|\lesssim\sqrt{m_*}$. Using this Gaussian approximation, we estimate the contribution of the region around $m=m_*$ to $Z_n$ as $\sqrt{2\pi m_*}g_{nm_*}q^{m_*}=e^{m_*}/n!$, the latter equation following from (\ref{pmstar}). The full estimate of $Z_n$, summing the contributions from the $m=0$ and $m=m_*$ maxima, is thus
\beq\label{Zn}
Z_n\approx 1+\frac{e^{m_*}}{n!}.
\eeq
The two terms in this approximation are equal if
\beq\label{mcr}
m_*=\log n!\approx n\log n = 2m_c,\quad
q=q_c\equiv\frac{2 \log n!}{n^2}.
\eeq
If $m_*$ is far away from $2m_c$ (or $q$ from $q_c$), then one of the two terms is negligible. Note that $2m_c$ lies within the validity region of the asymptotic approximation (\ref{lesssparse}) that we have used in our calculations. 

Next, using the approximations (\ref{expbe},\ref{pmstar},\ref{Zn}), we estimate the expected number of edges
\beq\label{mav}
\ea{m}\equiv \sum_m m\,p_{nm}\approx\frac{m_*}{1+n!/e^{m_*}},
\eeq
so at the critical point $m_*=2m_c$~\eqref{mcr}, the expected number of edges and the average degree are
\beq
 \ea{m}_c \approx \frac{m_*}{2} \approx m_c,\qquad
 \ea{k}_c \equiv \frac{2\ea{m}_c}{n}\approx k_c,\label{kavc}
\eeq
coinciding with the connectivity and symmetry transition in the labeled ER graphs at $k_c=\log n$. By $\ea{\cdots}_c$ we mean ensemble averages evaluated at $q=q_c$ of~(\ref{mcr}). We can also estimate the variance of $m$ as \cite{varfoot}
\beq
\Var(m)\approx\frac{m_*^2\, n!/e^{m_*}}{(1+n!/e^{m_*})^2}+\frac{2m_*}{1+n!/e^{m_*}}.
\eeq
 
\begin{figure}
  \centerline{\includegraphics[width=3.45in]{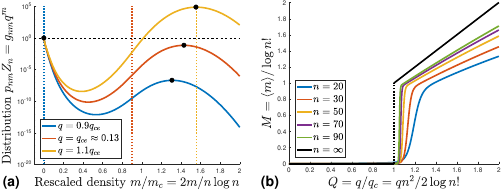}}
  \caption{{\it Broken equivalence and phase transitions:} \textbf{(a)}~The middle (red) curve shows the unnormalized distribution $p_{nm}Z_n$ for $n=50$ and the value of $q=q_{ce}\approx0.13$, which is such that the expected number of links $\ea{m}=\sum_m mp_{nm}\approx88$ matches the critical point maximizing the entropy derivative for the same $n$ in Fig.~\ref{figmicroen}(b) ($\ea{m}/m_c=m/m_c\approx0.90$). The other two curves show the same distribution for two values of $q$ slightly above and below this critical value. The vertical dotted lines indicate the corresponding values of~$\ea{m}$, while the black dots are the local maxima of the distributions. \textbf{(b)}~The rescaled values of $\ea{m}$ are shown as functions of the rescaled Boltzmann factor~$q$ for different values of~$n$, along with the theoretical thermodynamic limit in~\eqref{Minf}.\vspace{-3mm}}
  \label{fig2max}
\end{figure}

Imagine now we decrease $q$, as illustrated in Fig.~\ref{fig2max}(a). As long as $q\gg q_c$, the second term in (\ref{Zn}) dominates, so $\langle m \rangle\approx m_*$ and $\sigma(m)\equiv\sqrt{\Var(m)}\sim\sqrt{m_*}$, meaning that the high-probability values of $m$ are concentrated around $m_*$, and the canonical ensemble is identical at large $n$ to the microcanonical ensemble with $m=m_*$. On the other hand, if $q\ll q_c$, the first term in (\ref{Zn}) dominates, so $\langle m \rangle\approx 0$ and $\sigma(m)\approx 0$, and we are left with empty graphs. If we want the expected number of edges $\ea{m}$ in our canonical ensemble to be above $0$ but below $m_c$---i.e., if we want $0\neq\ea{{k}}\ll k_c=\log n$---then this is possible only for finite $n$, and only in a narrow region of $q$ slightly below $q_c$. In this case, both $\ea{m}$ and $\sigma(m)$ are $\sim m_*$, and the number of edges $m$ does not concentrate on a single value, since the two maxima of $p_{nm}$ at $m=0$ and $m=m_*$ provide comparable contributions. The graphs that contribute to the statistical average $\ea{m}$ are either empty, the probability of which in view of (\ref{Zn}-\ref{kavc}) is
\beq
P=p_{n0}=\frac{1}{Z_n}\approx1-\frac{\ea{m}}{m_*}\approx1-\frac{\ea{{k}}}{2k_c}\to1,
\eeq
or, with probability $1-P$, they have a number of edges close to $m_*$. None of these graphs have the number of edges close to $\langle m\rangle<m_*$, see Fig.~\ref{fig2max}(a). The equivalence with the microcanonical ensemble is thus broken \cite{grav}.

This observation is consistent with another notion of ensemble equivalence based on the large deviation principle~\cite{touchette}: two ensembles are equivalent if their specific relative entropy tends to zero:
\beq
s\equiv\frac{D_\mathrm{KL}}{n}\to0,
\eeq
where $D_\mathrm{KL}$ is the relative entropy, also known as the Kullback–Leibler divergence, between the microcanonical and canonical ensemble distributions. This can be seen as another form of the requirement that the canonical distribution with mean~$\ea{m}$ must be concentrated at the microcanonical value $m=\ea{m}$. In view of the observations above, this is definitely not the case here, so we expect $s$ to diverge.

To confirm this, we recall~\cite{squartini2015breaking} that $D_\mathrm{KL}=-\log p_{n,\ea{m}}$, which we estimate using (\ref{pmstar},\ref{gauss},\ref{mav}) as
\beq\label{pnmav}
 p_{n,\ea{m}} \approx \frac{\exp(-x_*^2/2m_*)}{\sqrt{2\pi m_*}(1+n!/e^{m_*})}, 
\eeq
with  $x_* = m_*-\ea{m}\approx{m_*}/(1+e^{m_*}/n!)$.
Next, we observe from~\eqref{mav} that if $m_*\sim n$, then $\ea{m}\ll1$, and if $m_*\sim n \log n$, then $\ea{m}\sim n \log n$. This implies that if we want our $\ea{m}$ to be such that $1\ll\ea{m}\ll n \log n$---e.g., $\ea{m}\sim n$ corresponding to a bounded average degree $\ea{{k}}\sim1$---then $m_*$ as a function of $n$ must satisfy $n \ll m_* \ll n \log n$. If so, \eqref{pnmav} implies that $-\log p_{n,\ea{m}} \sim m_*$, so $s\gg1$.

{\it First-order phase transition.}--- The situation described above is typical of first-order phase transitions~\cite{ident}: the probability is only supported on empty graphs and graphs with average degrees ${k}\sim\log n$, while attempting to tune the average degree to $\ea{{k}}\ll\log n$ results in a statistical mixture of these empty and logarithmic phases, and not in individual graphs with the desired average degree. The resulting abrupt change of the expected average degree $\ea{{k}}$ as a function of $q$ is seen in Fig.~\ref{fig2max}(b).

To show that this phase transition is indeed first-order, we analyze its thermodynamic limit by introducing new variables:
\beq
Q\equiv\frac{q}{q_c}=\frac{q\,n^2}{2 \log n!},\qquad M\equiv\frac{\ea{m}}{2m_c}=\frac{\langle m\rangle}{\log n!}.
\eeq
In these variables, $m_*\approx Q\log n!$, and (\ref{mav}) becomes
\beq\label{Mfiniten}
M(Q)\approx\frac{Q}{1+(n!)^{1-Q}}.
\eeq
This expression has an obvious $n\to\infty$ limit:
\beq\label{Minf}
M(Q)\to\begin{cases}\,\,Q&\mbox{if}\hspace{3mm}Q>1\hspace{3mm}(q>q_c),\\
\,\,0&\mbox{if}\hspace{3mm}Q<1\hspace{3mm}(q<q_c),
\end{cases}
\eeq
indicating a discontinuity of height 1 at $Q=1$ ($q=q_c)$, shown in Fig.~\ref{fig2max}(b) as well.

\vspace{5mm}

Our result that the unlabeled canonical \ER model in the sparse regime is not equivalent to its microcanonical cousin~\cite{Luczak} is sad news. If the two models were equivalent, the known properties of the microcanonical model, including its degree distribution and clustering~\cite{Luczak}, would characterize the canonical model as well, and would not have to be studied ``from scratch.'' Unfortunately, this is not the case, so even the degree distribution in the canonical model remains unknown, an interesting open question for future work.
 
In labeled graphs, statistical mixtures of sparser and denser phases are rather common~\cite{PN,PG,horvat2015reducing} and have been compared to gas-liquid phase transitions~\cite{liquid}. They occur, however, in more sophisticated ensembles that control, in addition to the number of edges, the number of triangles or other small subgraphs~\cite{RS,kenyon}. It is remarkable that, in unlabeled graphs, condensation phenomena of this kind are seen even in the simplest possible ensemble of random graphs---the \ER graphs---formulated in terms of the number of edges alone.

The fact that the phase transition in unlabeled graphs is at the same point $k_c=\log n$ as the graph symmetry transition in labeled graphs~\cite{ER2,cameron2013random} is not coincidental. Compared to a maximum-entropy distribution over labeled graphs, the corresponding distribution over unlabeled graphs can be seen as shifting the probability masses from more asymmetric to more symmetric graphs, thus making more symmetric graphs more likely~\cite{unlabeled}. But then, as soon as the graph density is lowered from high values, where almost all graphs are asymmetric~\cite{cameron2013random}, to the point $k_c=\log n$ that allows graphs to be symmetric, also allowing for the presence of nodes of degree~$0$, what is the simplest way to accommodate the shifted probability masses? As our results indicate, the simplest, maximum-entropy way to do so is just to move these masses to empty graphs, simply because they are maximally symmetric. In the microcanonical ensemble, the corresponding solution is to have a needed fraction of $0$-degree nodes---``poor loners'' in the social network language---since they contribute hyperexponentially to the amount of graph symmetry. It is remarkable that as soon as the expected average degree is $k \ll k_c$, these loners dominate in the thermodynamic limit of both the microcanonical and canonical ensembles.

It would be interesting to see if similar critical phenomena (related to more symmetric graphs being more likely) are present in more sophisticated/realistic unlabeled network models, including the stochastic block model, latent space models, and the configuration model \cite{config,configlbl}. Nothing is known about the unlabeled stochastic block model. First results concerning latent space models were obtained in \cite{unlabeled}, showing that sparse labeled and unlabeled one-dimensional random geometric graphs are entirely different even in terms of their entropy scaling. The existing results concerning the unlabeled configuration model say that if all nodes have the same degree~\cite{reg}, or even if the degree distribution is broad but the fractions of high- and low-degree nodes are sufficiently small~\cite{mckay1984automorphisms,brick2023threshold}, then there are no new interesting critical phenomena: the graphs are asymmetric, so the entropy of unlabeled graphs is the entropy of labeled graphs minus $\log n!$. In contrast, a ubiquitous feature of many real-world networks are large fractions of high-degree hubs and low-degree nodes~\cite{barabasi2016network,newman2018networks}, bolstering the network symmetry since star graphs are second to empty graphs in terms of the amount of symmetry. Does this feature of real-world networks lead to critical phenomena akin to those we have established here?

%%%%%%%%%%%%%%%%%%%%%%%%%%%%%%%%%
\begin{acknowledgments}
\rule{0mm}{10mm}We thank Omer Angel for interesting discussions and suggestions. OE is supported by Thailand NSRF via PMU-B (grant number B13F670063). DK is supported by NSF grant numbers CCF-2311160 and IIS-1741355.\vspace{2mm}
\end{acknowledgments}

\onecolumngrid

\onecolumngrid

\newpage

\begin{center}
{\bf\Large SUPPLEMENTAL MATERIAL}\vspace{7mm}
\end{center}
\twocolumngrid

\section*{Labeled vs.\ unlabeled network models and their applications}

Our focus in the main text has been on presenting surprising thermodynamic features of the simplest possible canonical ensemble of unlabeled graphs.
This problem evidently had to be considered before any other unlabeled graph ensembles. If more sophisticated and more realistic ensembles are studied,
it will be necessary to keep in mind the features of the simplest ensemble we have analyzed. It is interesting, however, to contemplate potential future applications.

We have remarked in the main text on possible connections to physical and material networks. A majority of models of such networks discussed in the past literature \cite{glover2024measuring,dehmamy2018structural,posfai2023impact,liu2020isotopy,pete2024physical,papadopoulos2018network,nauer2021random} are actually labeled. This is because most of them, but not all, are ``mechanistic,'' nonequilibrium models, as opposed to null models, e.g., standard canonical ensembles in equilibrium statistical mechanics. Classic examples of mechanistic models include models of absorption and preferential attachment. They tend to describe growing systems in which new elements join the system one at a time, and hence there is a preferred and physically meaningful labeling of elements by their birth times, or more generally, by some other system-specific variables.

The equilibrium null models in statistical mechanics are of very different nature. They do not attempt to reproduce or describe any physical dynamics of particles in the system. Instead, they specify the probability to find the system in a particular microscopic state, given all the macroscopic information about the system. All of the standard null models in equilibrium statistical mechanics, including the microcanonical, canonical, and grand canonical ensembles, can be derived from the maximum entropy principle~\cite{jaynes1957information}, which simply postulates that the probability distributions over microstates in the ensembles maximize the Shannon entropy of the distribution subject to the constrains coming from the macroscopic information about the system. This means that these models are actually \emph{defined} by direct specification of probability distributions over microstates. In network models, these microstates are individual graphs, and the macroscopic information is, for instance, the number of edges in the \ER model.

The labeled \ER model admits both interpretations. Mechanistically, it is a model in which ${n\choose2}$ distinguishable pairs of $n$ distinguishable elements are connected independently with probability~$p$. This can be formulated as a growth model, say, in terms of adding new edges one at a time. Yet, the model is also a standard canonical ensemble \emph{defined} as the probability distribution (over the space of labeled graphs on $n$ vertices) that maximizes the Shannon entropy, subject to the constraint that the average number of edges in these graphs is ${n\choose2}p$~\cite{park2004statistical}.

If the microscopic elements are now indistinguishable, like molecules and atoms in physical and material networks, then we believe it is obvious that null models associated with such networks must be models of unlabeled graphs. In case of the \ER model discussed here, its unlabeled version is still \emph{defined} as an entropy-maximizing distribution, except over unlabeled graphs, but as opposed to the labeled \ER model, its unlabeled version does not immediately suggest a mechanistic interpretation.

We note that using labeled null models to analyze networks with unlabeled nodes would lead to paradoxes similar to the (in)famous Gibbs paradox: treating indistinguishable particles as distinguishable leads to an incorrect, nonextensive expression for the entropy of the ideal gas. See~\cite{jaynes1992gibbs,swendsen} and references therein for the coverage of the dramatic 150-year-long history of the Gibbs paradox, and the role that quantum mechanics does (not!) play there.

In applications to physical networks, the unlabeled \ER model is probably not too realistic. One can imagine, in principle, disordered systems of molecules or atoms in which the expected number of connections---contacts or other interactions---is controlled by some external parameters like pressure. If such a system is described by the \ER model, then in view of the results presented here, the difference between labeled and unlabeled behaviors is drastic and can be easily detected in experiments. Indeed, as the pressure parameter is lowered from above to the critical point corresponding to the average degree $k_c = \log n$, nothing horrible happens to the labeled system: such a ``gel'' consisting of distinguishable particles acquires some small disconnected components, but the giant component is still there. If the particles are indistinguishable, however, then the gel ``evaporates''---the corresponding graph is empty with high probability.

We do not have an obvious example of a real-world system of particles for which the \ER model would be an appropriate null model, since particles in such systems exist in space, and spatial constraints play an important role. That is why a natural first step to make the unlabeled \ER model more realistic in application to physical networks is to make it spatial, e.g., by considering \emph{random geometric graphs} (RGGs), mentioned in~\cite{dehmamy2018structural,papadopoulos2018network,nauer2021random}. We reiterate, however, that these references deal with labeled RGGs, without mentioning potential differences with unlabeled RGGs, and that it was shown in~\cite{unlabeled} that even such a basic thermodynamic characteristic as the leading term of entropy is different in labeled versus unlabeled one-dimensional RGGs. It would be interesting to see these or similar differences confirmed in real-world experiments with physical networks, and our considerations of the simple unlabeled \ER ensemble are an essential prerequisite for such future studies.

\section*{P\'olya counting theory}

For the convenience of the readers, we outline here the calculations in the spirit of P\'olya counting theory that lead to the explicit form of the generating function for counting the unlabeled graphs:
\vspace{-2mm}
\beq
g_n(x)\equiv\sum_{m=0}^N g_{nm}x^m,
\eeq
where $g_{nm}$ is the number of unlabeled graphs on $n$ vertices with $m$ edges.
We use this expression in practice to obtain the graph counting data that underlie our numerics.
A systematic mathematical exposition and further details can be found in~\cite{HP}.

We first express $g_n(x)$ as a sum over all labeled graphs $G$ on $n$ vertices, but if we do so, we must divide the contribution of $G$ by the number of distinct labeled graphs generated from $G$ by label permutations:
\beq\label{aut}
g_n(x)=\sum_G\frac{x^{m(G)}}{\mbox{\# of distinct permutations of $G$}}. 
\eeq
The number of distinct permutations of $G$ can be represented as the total number of permutations $n!$, divided by the number of permutations that leave $G$ invariant. Correspondingly, (\ref{aut}) can be equivalently recast as
\beq\label{dePG}
g_n(x)=\frac1{n!}\sum_P\sum_G\de[G,P(G)]x^{m(G)}, 
\eeq
where $\sum_P$ is the sum over all permutations of the $n$ vertices, $P(G)$ is the result of applying the vertex permutation $P$ to $G$, and $\de[G,P(G)]$ equals 1 if $G=P(G)$ and 0 otherwise. To compute (\ref{dePG}), one appeals to the cycle decomposition of permutations: each permutation P can be represented by splitting the vertices into groups: $j_1$ groups with 1 vertex each, $j_2$ groups with 2 vertices each, and so on (clearly, $j_1+2j_2+3j_3+\cdots=n$), and making $P$ act as a cyclic shift by one unit $1\to 2\to 3\to\cdots\to 1$ within each group. There are $n!/(\prod_k k^{j_k}j_k!)$ ways to assign specific vertices to a given cycle structure $\{j_k\}$. Since (\ref{dePG}) is invariant under vertex renumbering, this lets one reduce the sum over all permutations in (\ref{dePG}) to a sum over all cycle structures $\{j_k\}$:
\beq
g_n(x)=\sum_{\{j_k\}}\Big(\prod_k k^{j_k}j_k!\Big)^{\hspace{-1mm}-1}\hspace{-2mm}\sum_{G\,\,\mathrm{inv}\,P_{\{j_k\}}}\hspace{-3mm}x^{m(G)}, 
\eeq
where $P_{\{j_k\}}$ is any representative permutation with the cycle structure given by $j_k$, $\sum_{G\,\,\mathrm{inv}\,P_{\{j_k\}}}$ denotes summation over all graphs left invariant by $P_{\{j_k\}}$, and $\sum_{\{j_k\}}$ denotes summation over all $j_k$ under the condition $\sum_k kj_k=n$. Finally, each vertex permutation $P$ induces the corresponding pair-permutation $\mathcal{P}$ on pairs of vertices, which are potential edges that may be filled or not filled. The pair-permutation $\mathcal{P}$ has its own cycle structure, splitting the set of would-be edges into groups, with the edges of each group permuted cyclically. The graph is only invariant under $P$ if, for each cycle of the corresponding pair-permutation $\mathcal{P}$ independently, the edges are either all empty, giving a factor 1 in $x^{m(G)}$, or all filled, giving a factor of $x^{\mbox{\tiny cycle length}}$. Hence, schematically,
\beq\label{cyclen}
g_n(x)=\sum_{\{j_k\}}\Big(\prod_k k^{j_k}j_k!\Big)^{\hspace{-1mm}-1}\hspace{-5mm}\prod_{\mbox{\tiny cycles of }\mathcal{P}_{\{j_k\}}}\hspace{-3mm}\left(1+x^{\mbox{\tiny cycle length}}\right), 
\eeq
What remains is to compute, for each choice of $j_k$, the number of cycles of each length in the pair-permutation $\mathcal{P}_{\{j_k\}}$ induced by the vertex permutation $P_{\{j_k\}}$. This computation is performed in \cite{HP}, resulting in the expression quoted in the main text:
\begin{align*}
g_n(x)=&\sum_{\{j_k\}}\Big(\prod_k k^{j_k}j_k!\Big)^{\hspace{-1mm}-1}\prod_{r<s} (1+x^{[r,s]})^{(r,s)j_rj_s}\\
&\times\prod_k \Big\{(1+x^{2k+1})^{kj_{2k+1}}(1+x^{2k})^{(k-1)j_{2k}}\\[-1mm]
&\hspace{1.3cm}\times (1+x^k)^{j_{2k}+kj_k(j_k-1)/2}\Big\},
\end{align*}
where $[r,s]$ and $(r,s)$ are the least common multiple and the greatest common divisor of $r$ and $s$, while $\sum_{\{j_k\}}$ stands for summation over all nonnegative integers $\{j_k\}$ such that $\sum_{k=1}^nkj_k=n$.


\begin{thebibliography}{99}
\twocolumngrid

\bibitem{erdos1959random}P.~Erd\H{o}s and A.~R\'enyi, {\it On random graphs I}, Pub.\ Math.\ Debrecen {\bf 6} (1959) 290. 

\bibitem{bollobas1985random}B.~Bollob{\'{a}}s, {\it \doi{Random graphs}{10.1017/CBO9780511814068}} (Academic Press, 1985).
\bibitem{janson2000random}S.~Janson, T.~{\L}uczak and A.~Rucinski, {\it \doi{Random graphs}{10.1002/9781118032718}} (Wiley, 2000).
\bibitem{hofstad2016random}R.~van~der~Hofstad, {\it \doi{Random graphs and complex\\ networks}{10.1017/9781316779422}} (Cambridge, 2016).
\bibitem{barabasi2016network}A.-L.~Barab{\'{a}}si, {\it \href{http://www.cambridge.org/9781107076266}{Network science}} (Cambridge, 2016).
\bibitem{newman2018networks}M.~Newman, {\it \doi{Networks: an introduction}{10.1093/oso/9780198805090.001.0001}} (Oxford, 2018).

\bibitem{park2004statistical}J.~Park and M.~E.~J.~Newman, {\it Statistical mechanics of networks,} \doi{Phys.\ Rev.\ E {\bf 70} (2004) 066117}{10.1103/PhysRevE.70.066117}, \arXiv{cond-mat/0405566}.

\bibitem{contraction}The same ensemble also emerges in contracting networks under various random deletion scenarios, see I.~Tishby, O.~Biham and E.~Katzav, {\it Convergence towards an \ER graph structure in network contraction processes,} \doi{Phys.
Rev. E {\bf 100} (2019) 032314}{10.1103/PhysRevE.100.032314}, \arXiv{2009.02890} [physics.soc-ph];
{\it Analysis of the convergence of
the degree distribution of contracting random networks towards a
Poisson distribution using the relative entropy,} \doi{Phys. Rev. E {\bf 101} (2020) 062308}{10.1103/PhysRevE.101.062308}, \arXiv{2009.04249} [physics.soc-ph].

\bibitem{anand2010gibbs}K.~Anand and G.~Bianconi, {\it Gibbs entropy of network ensembles by cavity methods,} \doi{Phys.\ Rev.\ E {\bf82} (2010) 011116}{10.1103/PhysRevE.82.011116}, \arXiv{1001.4717} [cond-mat.dis-nn].

\bibitem{squartini2015breaking}T.~Squartini, J.~de Mol, F.~den Hollander and D.~Garlaschelli,
{\it Breaking of ensemble equivalence in networks}, \doi{Phys. Rev. Lett. {\bf 115} (2015) 268701}{10.1103/PhysRevLett.115.268701}, \arXiv{1501.00388} [cond-mat.stat-mech].

\bibitem{ER1}P.~Erd\H{o}s and A.~R\'enyi, {\it On the evolution of random graphs},
\doi{Mag.\ Tud.\ Akad.\ Mat.\ Kut.\ Int.\ K\H{o}z. {\bf 5} (1960) 17}{10.1515/9781400841356.38}.

\bibitem{aldous1997brownian}D.~Aldous, {\it Brownian excursions, critical random graphs and the multiplicative coalescent}, \doi{Ann.\ Prob.\ {\bf 25} (1997) 812}{10.1214/aop/1024404421}.

\bibitem{ER2}P.~Erd\H{o}s and A.~R\'enyi, {\it Asymmetric graphs,} 
\doi{Acta Math.\ Hung. {\bf 14} (1963) 295}{10.1007/BF01895716}.

\bibitem{cameron2013random}P.~J.~Cameron, {\it The random graph} in {\it Math Paul Erd\H{o}s II} (2013), \doi{pp.~353-378}{10.1007/978-1-4614-7254-4_22}.

\bibitem{symmetry}A labeled graph is called \emph{asymmetric} if any permutation of its node labels leads to a different (but isomorphic) labeled graph. Otherwise, the graph is called \emph{symmetric}. The labeled \ER graph is asymmetric with high probability if $k\gg\log n$, and symmetric with high probability if $k\ll\log n$~\cite{ER2,cameron2013random}.

\bibitem{touchette}H.~Touchette, {\it Equivalence and nonequivalence of ensembles: thermodynamic, macrostate, and measure levels}, 
\doi{J.\ Stat.\ Phys. {\bf 159} (2015) 987}{10.1007/s10955-015-1212-2}, \arXiv{1403.6608} [cond-mat.stat-mech];
 {\it Asymptotic equivalence of probability measures and stochastic processes},
\doi{J.\ Stat.\ Phys. {\bf 170} (2018) 962}{10.1007/s10955-018-1965-5}, \arXiv{1708.02890} [cond-mat.stat-mech].

\bibitem{Luczak}T.~\L uczak, {\it How to deal with unlabeled random graphs},\\
\doi{J.\ Graph\ Th. {\bf 15} (1991) 303}{10.1002/jgt.3190150307}.

\bibitem{papadopoulos2018network}L.~Papadopoulos, M.~A.~Porter, K.~E.~Daniels and D.~S.~Bassett, {\it Network analysis of particles and grains,}
\doi{J.\ Compl.\ Net. {\bf 6} (2018) 485}{10.1093/comnet/cny005}, \arXiv{1708.08080} [cond-mat.soft].

\bibitem{dehmamy2018structural}N.~Dehmamy, S.~Milanlouei and A.-L.~Barab\'asi, {\it A structural transition in physical networks,} \doi{Nature {\bf 563} (2018) 676}{10.1038/s41586-018-0726-6}.

\bibitem{nauer2021random}S.~Nauer, L.~B\"ottcher and M.~A.~Porter, {\it Random-graph models and characterization of granular networks,} 
\doi{J.\ Compl.\ Net. {\bf 8} (2020) cnz037}{10.1093/comnet/cnz037}, \arXiv{1907.13424} [cond-mat.soft].

\bibitem{liu2020isotopy}Y.~Liu, N.~Dehmamy and A.-L.~Barab\'asi, {\it Isotopy and energy of physical networks,} \doi{Nature Phys. {\bf 17} (2021) 216}{10.1038/s41567-020-1029-z}.

\bibitem{posfai2023impact}M.~P\'osfai, B.~Szegedy, I.~Ba\v{c}i\'c, L.~Blagojevi\'c, M.~Ab\'ert, J.~Kert\'esz, L.~Lov\'asz and A.-L.~Barab\'asi, {\it Impact of physicality on network structure,} \doi{Nature Phys. {\bf 20} (2024) 142}{10.1038/s41567-023-02267-1}, \arXiv{2211.13265} [cond-mat.stat-mech].

\bibitem{pete2024physical}G.~Pete, \'A.~Tim\'ar, S.~\"O.~Stef\'ansson, I.~Bonamassa and M.~P\'osfai, {\it Physical networks as network-of-networks,} \doi{Nature Comm. {\bf 15} (2024) 4882}{10.1038/s41467-024-49227-8}, \arXiv{2306.01583} [cond-mat.stat-mech].

\bibitem{glover2024measuring}C.~Glover and A.-L.~Barab\'asi, {\it Measuring entanglement in physical networks}, \doi{Phys.\ Rev.\ Lett.\ {\bf 133} (2024) 077401}{10.1103/PhysRevLett.133.077401}, \arXiv{2403.01328} [cond-mat.dis-nn].

\bibitem{quantum}In the Supplemental Material that further contains references \cite{jaynes1957information,jaynes1992gibbs,swendsen}, we provide additional comments on real-world applications of unlabeled network ensembles, along with a brief summary of unlabeled graph combinatorics.

\bibitem{jaynes1957information}E.~T.~Jaynes, {\it Information theory and statistical mechanics,} \doi{Phys. Rev. {\bf 106} (1957) 620}{10.1103/PhysRev.106.620}.

\bibitem{jaynes1992gibbs}E.~T.~Jaynes, {\it The Gibbs paradox}, in \doi{Maximum entropy and Bayesian methods}{10.1007/978-94-017-2219-3_1} (Springer, 1992).

\bibitem{swendsen}R.~H.~Swendsen, { \it Statistical mechanics of classical systems with distinguishable particles,} 
\doi{J.\ Stat.\ Phys. {\bf 107} (2002) 1143}{10.1023/A:1015161825292}; {\it Probability, entropy, and Gibbs' paradox(es),}
\doi{Entropy {\bf 20} (2018) 450}{10.3390/e20060450}.

\bibitem{wormald}N.~C.~Wormald, {\it Generating random unlabelled graphs,} \doi{SIAM\ J.\ Comp. {\bf 16} (1987) 717}{10.1137/0216048}.

\bibitem{canonical}B.~D.~McKay and A.~Piperno, {\it Practical graph isomorphism II,} \doi{J.\ Symb.\ Comp. {\bf 60} (2014) 94}{10.1016/j.jsc.2013.09.003}, \arXiv{1301.1493} [cs.DM].

\bibitem{gross2001microcanonical}D.~H.~E.~Gross, {\it \doi{Microcanonical thermodynamics}{10.1142/4340}}
(World Scientific, 2001).

\bibitem{nonconc}H.~Behringer and M.~Pleimling,
{\it Continuous phase transitions with a convex dip in the microcanonical entropy},
\doi{Phys. Rev. E {\bf 74} (2006) 011108}{10.1103/PhysRevE.74.011108},
\arXiv{cond-mat/0606283}.

\bibitem{horvat2015reducing}Sz.~Horv{\'{a}}t, \'E.~Czabarka and Z.~Toroczkai, {Reducing degeneracy in maximum entropy models of networks,}
\doi{Phys.\ Rev.\ Lett. {\bf 114} (2015) 158701}{10.1103/PhysRevLett.114.158701}, \arXiv{1407.0991} [cond-mat.stat-mech].

\bibitem{thesis}B.~W.~Stortenbecker, {\it The entropic concavity framework: a universal formulation for understanding phase transitions,}
\href{https://www.proquest.com/openview/dbf288881fcc53c681f27601b75cbb46/}{PhD thesis} at the University of Notre Dame (2022).

\bibitem{OEIS}OEIS sequence \href{https://oeis.org/A008406}{A008406}.

\bibitem{geo}There is some similarity between vertex renumbering and choosing a coordinate system in continuous Riemannian spaces. Contacts between random graphs and discretized spatial geometries have been made in\\
S.~Chen and S.~S.~Plotkin,
{\it Statistical mechanics of graph models and their implications for emergent spacetime manifolds,}
\doi{Phys. Rev. D \textbf{87} (2013) 084011}{10.1103/PhysRevD.87.084011}, \arXiv{1210.3372} [gr-qc];\\
C.~A.~Trugenberger,
{\it Combinatorial quantum gravity: geometry from random bits,}
\doi{JHEP \textbf{09} (2017) 045}{10.1007/JHEP09(2017)045}, \arXiv{1610.05934} [hep-th];\\
P.~Akara-pipattana, T.~Chotibut and O.~Evnin,
{\it The birth of geometry in exponential random graphs,}
\doi{J.\ Phys.\ A \textbf{54} (2021) 425001}{10.1088/1751-8121/ac2474},
\arXiv{2102.11477} [cond-mat.dis-nn];\\
P.~van der Hoorn, W.~J.~Cunningham, G.~Lippner, C.~Trugenberger and D.~Krioukov,
{\it Ollivier-Ricci curvature convergence in random geometric graphs,}
\doi{Phys. Rev. Res. \textbf{3} (2021) 013211}{10.1103/PhysRevResearch.3.013211}, \arXiv{2008.01209} [math.PR].

\bibitem{HP}F.~Harary and E.~M.~Palmer, {\it\doi{Graphical enumeration}{10.1016/C2013-0-10826-4}} (Academic Press, 1973), chapter 4.

\bibitem{num} The number of terms in the sum over $j_k$ in (\ref{Harary}) grows like the number of integer partitions of $n$, which is $\exp[{O(\sqrt{n})]}$ at large $n$. While this superpolynomial growth does create difficulties in numerically implementing (\ref{Harary}) for large graphs, the problem is evidently much simpler than going over either all $n!$ permutations or all $2^{N}$ labeled graphs.

\bibitem{share} The code and data are available at \url{http://bitbucket.org/dk-lab/2024_code_data_unlabeled}.

\bibitem{Wright}E.~M.~Wright, {\it Graphs on unlabelled nodes with a large number of edges,}
\doi{Proc. London Math. Soc. {\bf 28} (1974) 577}{10.1112/plms/s3-28.4.577}.

\bibitem{unlabeled}J.~Paton, H.~Hartle, H.~Stepanyants, P.~van der Hoorn and D.~Krioukov,
{\it Entropy of labeled versus unlabeled networks}, \doi{Phys. Rev. E {\bf 106} (2022) 054308}{10.1103/PhysRevE.106.054308
}, \arXiv{2204.08508} [physics.soc-ph].

\bibitem{labelmax}For labeled graphs, (\ref{pmstar}) would have been given by the same formula, but without $n!$ in the denominator. This changes the situation drastically: without $n!$, $e^{m_*}/\sqrt{2\pi m_*}$ is always much greater than 1 at $m_*\gg 1$, so the probability of $m=0$, given by $1/Z_n$, is negligible compared to the probability of $m=m_*$.

\bibitem{varfoot}For computing the variance, the spread $m$ around $m_*$ captured by the $x$-dependence in (\ref{gauss}) plays a role, to account for which, we simply perform the summation $\sum_x (m_*+x)^2 e^{-x^2/m_*}=\sqrt{2\pi m_*}(m_*^2+2m_*)$. Note that, as in (\ref{Zn}), the sum over $x$ can be approximated by an integral with an arbitrary precision when $m_*\gg 1$.

\bibitem{grav}It is interesting to contemplate the parallels between breaking ensemble equivalence due to indistinguishability of configurations related by permutations
and similar thermodynamic phenomena occuring in the presence of long-range interactions:\\
D.~Lynden-Bell, {\it Statistical mechanics of violent eelaxation in
stellar systems,} \doi{Mon.\ Not.\ Royal Astr.\ Soc.
{\bf 136} (1967) 101}{10.1093/mnras/136.1.101};\\
F.~Bouchet and J.~Barr\'e, {\it Classification of phase transitions and
ensemble inequivalence in systems with long range interactions,} 
\doi{J.\ Stat.\ Phys. {\bf 118} (2005) 1073}{10.1007/s10955-004-2059-0}, \arXiv{cond-mat/0303307};\\
 A.~Campa, A.~Giansanti, G.~Morigi and F.~Sylos Labini,
{\it Dynamics and thermodynamics of systems with long-range interactions:
theory and experiments,} AIP conference proceedings {\bf 970} (2008);\\
 A.~Campa, T.~Dauxois, D.~Fanelli and S.~Ruffo, {\it \doi{Physics of long-range interacting systems}{10.1093/acprof:oso/9780199581931.001.0001}} (Oxford, 2014).

\bibitem{ident}Similar phase-transition behaviors have been observed in other systems with enforced permutation equivalence:\\
J.-H.~Park and S.-W.~Kim,
{\it Thermodynamic instability and first-order phase transition in an ideal Bose gas,}
\doi{Phys. Rev. A {\bf 81} (2010) 063636}{10.1103/PhysRevA.81.063636},
\arXiv{0809.4652} [cond-mat.stat-mech];\\
D.~O'Connor and S.~Ramgoolam,
{\it Permutation invariant matrix quantum thermodynamics and negative specific heat capacities in large $N$ systems,} 
\doi{JHEP \textbf{12} (2024) 161}{10.1007/JHEP12(2024)161},
\arXiv{2405.13150} [hep-th].

\bibitem{PN} J.~Park and M.~E.~J.~Newman, {\it Solution of the two-star model of a network,} 
\doi{Phys.\ Rev.\ E {\bf 70} (2004) 066146}{10.1103/PhysRevE.70.066146}, \arXiv{cond-mat/0405457};
{\it Solution for the properties of a clustered network}, 
\doi{Phys.\ Rev.\ E {\bf 72} (2005) 026136}{10.1103/PhysRevE.72.026136}, \arXiv{cond-mat/0412579}.

\bibitem{PG}G.~Bizhani, M.~Paczuski and P.~Grassberger, {\it Discontinuous percolation transitions
in epidemic processes, surface depinning in random media, and Hamiltonian random
graphs}, \doi{Phys. Rev. E {\bf 86} (2012) 011128}{10.1103/PhysRevE.86.011128}, \arXiv{1202.3136} [cond-mat.dis-nn].

\bibitem{liquid}J.~Neeman, Ch.~Radin and L.~Sadun, {\it Existence of a symmetric bipodal phase in the
edge-triangle model,} \doi{J.\ Phys.\ A {\bf 57} (2024) 095003}{10.1088/1751-8121/ad259d}, \arXiv{2211.10498} [math.PR].

\bibitem{RS}Ch.~Radin and L.~Sadun, {\it Phase transitions in a complex network,}
\doi{J.\ Phys. A {\bf 46} (2013) 305002}{10.1088/1751-8113/46/30/305002}, \arXiv{1301.1256} [math-ph].

\bibitem{kenyon}R.~Kenyon, Ch.~Radin, K.~Ren and L.~Sadun, 
{\it Multipodal structure and phase transitions in large constrained graphs},
\doi{J.\ Stat.\ Phys. {\bf 168} (2017) 233}{10.1007/s10955-017-1804-0}, \arXiv{1405.0599} [math.CO];
{\it Bipodal structure in oversaturated random graphs},
\doi{Int.\ Math.\ Res.\ Not. {\bf 2018} (2018) 1009}{10.1093/imrn/rnw261}, \arXiv{1509.05370} [math.CO];
{\it The phases of large networks with edge and triangle constraints,}
\doi{J.\ Phys.\ A {\bf 50} (2017) 435001}{10.1088/1751-8121/aa8ce1}, \arXiv{1701.04444} [math.CO].

\bibitem{config}There is a series of recent works that demonstrate how to incorporate degree constraints into statistical physics computations:\\
T.~Kawamoto, {\it Entropy of microcanonical finite-graph ensembles,}
 \doi{J. Phys. Compl. {\bf 4} (2023) 035005}{10.1088/2632-072X/acf01c},
\arXiv{2305.10996} [cond-mat.stat-mech]; \\
O.~Evnin and W.~Horinouchi,
{\it A Gaussian integral that counts regular graphs,}
\doi{J. Math. Phys. \textbf{65} (2024) 093301}{10.1063/5.0208715},
\arXiv{2403.04242} [cond-mat.stat-mech];\\ 
P.~Akara-pipattana and O.~Evnin,
{\it Statistical field theory of random graphs with prescribed degrees,}
\arXiv{2410.11191} [cond-mat.stat-mech].\\
By blending such considerations with P\'olya counting theory, one may hope to bring under control the unlabeled configuration model.

\bibitem{configlbl}Entropy of labeled configuration model
networks has been treated in
A.~Annibale, A.~C.~C.~Coolen, L.~P.~Fernandes,
F.~Fraternali and J.~Kleinjung, {\it Tailored graph ensembles as proxies or
null models for real networks I: tools for quantifying structure,} 
\doi{J. Phys. A {\bf 42} (2009) 485001}{10.1088/1751-8113/42/48/485001}, \arXiv{0908.1759} [cond-mat.dis-nn].

\bibitem{reg}B.~Bollob\'as, {\it The asymptotic number of unlabelled regular graphs,} \doi{J.\ London\ Math.\ Sci. {\bf 26} (1982) 201}{10.1112/jlms/s2-26.2.201}.

\bibitem{mckay1984automorphisms}B.~D.~McKay and N.~C.~Wormald, {\it Automorphisms of random graphs with specified vertices,}
\doi{Combinatorica {\bf 4} (1984) 325}{10.1007/BF02579144}.

\bibitem{brick2023threshold}L.~Brick, P.~Gao and A.~Southwell, {\it The threshold of symmetry in random graphs with specified degree sequences,}
\doi{SIAM J.\ Discr.\ Math. {\bf 37} (2023) 94}{10.1137/21M1395296},
\arXiv{2004.01794} [math.CO].

\end{thebibliography}
\end{document}